\documentclass{IEEEtran}
\usepackage{cite, hyperref}
\usepackage{amsmath,amssymb,amsfonts}
\usepackage{algorithmic}
\usepackage{graphicx}
\usepackage{textcomp}
\usepackage{threeparttable}
\usepackage{multicol, blindtext,xcolor}
\usepackage{siunitx}
\usepackage[normalem]{ulem}
\usepackage{float}
\definecolor{darkgreen}{rgb}{0.0, 0.5, 0.0}
\def\BibTeX{{\rm B\kern-.05em{\sc i\kern-.025em b}\kern-.08em
    T\kern-.1667em\lower.7ex\hbox{E}\kern-.125emX}}
\begin{document}
\title{Semianalytically Designed Dual Polarized Printed-Circuit-Board (PCB) Metagratings}
\author{Yuval Shklarsh and Ariel Epstein, \IEEEmembership{Senior Member, IEEE}
\thanks{The authors are with the Andrew and Erna Viterbi Faculty of Electrical and Computer Engineering, Technion - Israel Institute of Technology, Haifa 3200003, Israel (e-mail: shklarsh@technion.ac.il; epsteina@ee.technion.ac.il).}
\thanks{Manuscript received XX,YY,2022; revised XX,YY, 2022.}}

\markboth{IEEE Transactions on Antennas and Propagation,~Vol.~XX, No.~YY, ZZ~2023}%
{Shklarsh and Epstein}

\maketitle

\begin{abstract}
Metagratings (MGs), sparse (periodic) composites of subwavelength polarizable particles (meta-atoms), have demonstrated highly efficient diffraction engineering capabilities via meticulous tailoring of the interaction between individual scatterers. To date, MGs at microwave frequencies have mostly been devised for either transverse electric (TE) or transverse magnetic (TM) polarized scenarios, which limits their use in many practical applications. Herein, we bridge this gap and present a comprehensive semianalytical design method for dual-polarized MGs with a separable response for TE and TM waves. First, by relating a printed circuit board (PCB) compatible array of dog-bone elements with the canonical dipole line analytical model, we establish a meta-atom for TM-polarized MGs, featuring negligible interaction with TE waves. Subsequently, we integrate the proposed configuration with a systematic synthesis scheme to implement a TM beam splitter MG, harnessing the equivalent dipole line model to resolve analytically the optimal meta-atom coordinates and dog-bone polarizability, without resorting to full-wave optimization. 
Finally, we show that a dual-polarized MG beam splitter can be conveniently synthesized correspondingly, combining the TM-polarized structure \emph{as is} with previously reported TE-polarized MG designs. This work paves a clear path towards integration of sparse, semianlaytically synthesized, efficient MGs in practical dual-polarized communication and imaging applications. 
\end{abstract}


\section{Introduction}
\label{Introduction}
\IEEEPARstart{I}{n} recent years, the emergence of semianalytical design tools for electromagnetic composites have enabled the creation of new advanced devices unavailable before. In such configurations, where the structure consists of many degrees of freedom (DOF), beyond the capability of the conventional full-wave optimizers, analytically-based models can be employed to reduce the computation time and yet reach near-optimal performance.
In the realm of metasurfaces, these ideas manifest themselves in the macroscopic design scheme \cite{tretyakov2003analytical, kuester2003averaged, holloway2012overview, epstein2016huygens, glybovski2016metasurfaces, hu2021review}. Metasurfaces, dense planar arrangements of subwavelength polarizable particles termed meta-atoms (MAs), are typically designed in two separate steps: first, the macroscopic design step, where a continuously varying surface with a different surface impedance in each location is manifested to reach a desired functionality; second, the microscopic design step, where a batch of MAs is designed in various methods, matching a suitable physical structure to every local response prescribed by the former step.
Transmissive beam deflectors\cite{lalanne1998blazed, bomzon2002space, yu2011light, pfeiffer2013metamaterial, monticone2013full, Selvanayagam2013, epstein2016arbitrary, Asadchy2016}, flat lenses\cite{yang2014efficient, lin2014dielectric, Asadchy2015}, and metasurface-based antennas\cite{epstein2016cavity, macifaenzi2019metasurface, boyarsky2021electronically}, all rely on analytical macroscopic design techniques, are only some examples to devices that have demonstrated unprecedented efficacy by utilizing the aforementioned combined synthesis schemes.

Nonetheless, the microscopic design step described above poses certain difficulties. First, it requires the implementation of many, densely-arranged, often geometrically complex, MAs. This could lead to significant fabrication and design challenges, and 
tends to involve time-consuming full-wave optimizations \cite{pfeiffer2014bianisotropic, epstein2016cavity, cole2018refraction, Chen2018, lavigne2018susceptibility}. Indeed, 
some simplifications can be achieved by considering cascaded impedance sheet models, utilized to devise Huygens' and bianisotropic meta-atoms\cite{monticone2013full, pfeiffer2014bianisotropic, epstein2016arbitrary}. Yet, these techniques still lack crucial aspects, neglecting interlayer near-field interactions and often ignoring conductor loss; consequently, final optimization in full-wave solvers is almost always necessary. Second, metasurface design procedures rely on the assumption that when the separately designed MAs are combined together, the abstract continuous response prescribed by the macroscopic design would be reproduced (the homogenization approximation). 
This assumption, however, is hard to justify rigorously, especially for general metasurfaces with highly-inhomogeneous polarizability distributions, 
since the intercoupling between different neighboring MAs is typically not considered in the microscopic design scheme. Furthermore, it is not \textit{a priori} clear how to determine the discretization resolution (meta-atom size) that would guarantee successful emulation of the homogenized response \cite{estakhri2016wave}. Therefore, relying on these approaches may make it difficult to identify the root cause of discrepancies between theoretical predictions and the ultimate metasurface performance, when such are observed. 

One of the developments that was motivated by these microscopic design challenges when applied to practical complex media devices was the concept of metagratings (MGs). These sparse (typically periodic) arrangements of subwavelength polarizable particles (MAs) \cite{ra2021metagratingsreview} have shown great potential in producing novel high efficiency beam-manipulating devices \cite{sell2017large, memarian2017wide, radi2017metagratings, wong2018perfect}. In contrast to metasurfaces, MGs are devised by considering the interactions and mutual coupling between the MAs using reliable analytical models \cite{radi2017metagratings, epstein2017unveiling}. These models, enabling delicate tunning of the interference patterns formed by the MGs upon external excitation, faciliated a large number of devices demonstrating a variety of wavefront manipulation functionalities \cite{ra2021metagratingsreview}.
Indeed, MGs were utilized across the electromagnetic spectrum to realize perfect anomalous reflection \cite{radi2017metagratings, wong2018perfect}, refraction \cite{sell2017large, yang2018freeform, dong2020efficient}, and focusing devices \cite{paniagua2018metalens, kang2020efficient}.

In particular, in recent years, we have developed and utilized such a suitable analytical model to realize printed circuit board (PCB) MGs that perform versatile diffraction engineering tasks at microwave frequencies, devising the complete fabrication-ready layout without
time consuming full-wave optimizations \cite{epstein2017unveiling, rabinovich2018analytical, rabinovich2019experimental, arbitrary2020}. Alas, as the model and subsequently the resultant MGs rely on elongated loaded wires as the underlying MAs\cite{epstein2017unveiling, tretyakov2003analytical}, this design scheme can only accommodate transverse electric (TE) polarized functionalities. Despite this alleged limitation, this convenient configuration and theoretical framework were further harnessed and elaborated by various groups to tackle a broad range of applications and scenarios, making it one of the more common embodiments for MGs at microwave frequencies to date \cite{popov2019beamforming, casolaro2019dynamic, popov2020conformal, xu2020dual, arbitrary2020, popov2021non, killamsetty2021metagratings, xu2021analysis, xu2022extreme, kerzhner2022metagratingsidelobssuppression, Liranwaveguiebends2022}.

Indeed, some researchers have suggested to address transverse magnetic (TM) polarized waves via MGs using vertical conducting loops \cite{popov2019designing}, in which induced currents emulate a magnetic line source response, or by using grooves in metallic slabs\cite{dualpolgrooves2020, rahmanzadeh2020perfect, rajabalipanah2021analytical, rahmanzadeh2022analysis, rajabalipanah2022parallel}, which support propagating TM modes. However, these solutions are not compatible with low-profile PCB technology, and may lead to challenging assembly and bulky devices. Another alternative considered for TM-polarized MGs is to utilize 
wide strip arrays (essnetially, narrow slot arrays), which interact with the orthogonal polarization following Babinet's principle \cite{Volakisstrip2004,stripsTMtretyakov2008, memarian2017wide,reconfigurable_strips2018, rahmanzadeh2021analytical}. Nonetheless, these strips also interact with TE polarized waves, making it more difficult to design a device with different responses to the two orthogonal polarizations.

\begin{figure}[tbp]
\centering
\includegraphics[width=80mm]{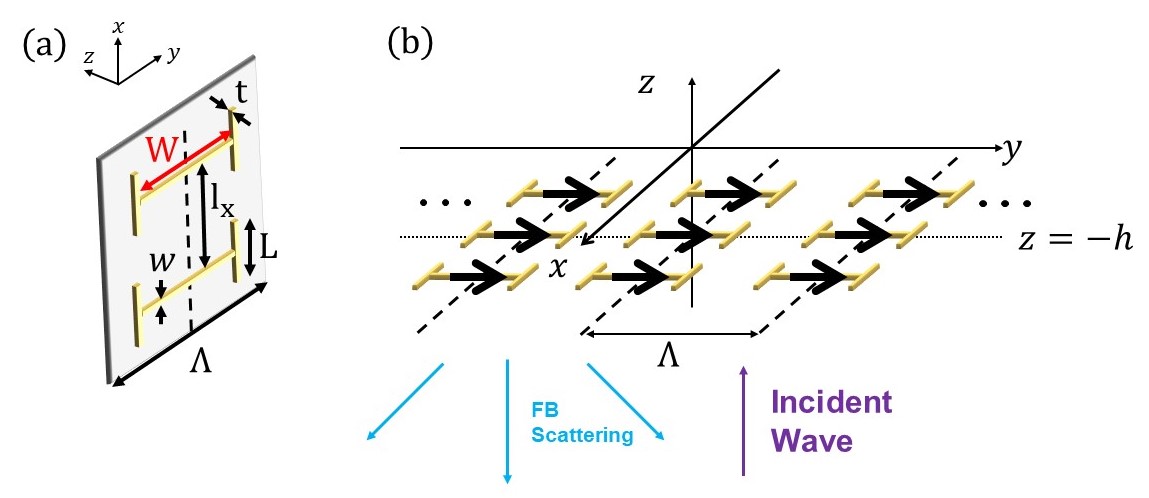}
\caption{Physical configuration of a TM-polarized MG positioned on the $xy$ plane. (a) Dog-bone shaped meta-atoms of width $W$, serving as the main design DOF, repeating with distance $l_x$ along the $x$ axis. (b) $\Lambda$-periodic MG positioned on plane $z=-h$, formed by the dog-bone meta-atom arrays. When a normally incident TM-polarized plane impinges on the structure, localized currents (modelled as point electric dipoles having dipole moment $\vec{p}$ - denoted by black arrows - corresponding to the leading term in the multipole expansion) would be induced on the dog-bones, causing scattering to a discrete set of angles according to the Floquet-Bloch (FB) theorem.}
\label{fig:configuration_fig}
\end{figure}

In this paper, we propose an alternative solution, developing a comprehensive semianalytical design scheme for dual-polarized PCB  compatible MGs. Specifically, we propose to utilize properly aligned 
dog-bone arrays as MAs \cite{tretyakov2000line}, which are susceptible to TM-polarized fields while exhibiting negligible response to TE-polarized excitations (Fig. \ref{fig:configuration_fig}). As shown, since the electric dipoles induced on this meta-atom configuration fit the canonical dipole line model \cite{felsenbook}, MG analysis and synthesis schemes can be conveniently augmented to analytically accommodate this new geometry \cite{shklarsh2021semianalytically}. Importantly, as the resultant TM-type MG does not interact with TE-polarized fields, it can be readily combined with the standard loaded-wire embodiment of TE-polarized MGs \cite{arbitrary2020} to create separable-response, dual-polarized, PCB MGs. 

We demonstrate our approach by designing a highly-efficient wide-angle polarization-dependent beam-splitter, combining independently devised TE- and TM- susceptible configurations to impose different splitting angles for each polarization. Contrary to metasurfaces previously developed to implement similar functionalities, incorporating dual-polarized control via dense Jerusalem crosses or crossed-dipole meta-atoms \cite{pfeiffer2013millimeter, selvanayagam2014polarization, cui2020dual}, the proposed MG-based devices are shown to offer a significant complexity reduction, featuring a sparse final layout and simplified design process.
As verified via full-wave simulations, the formulated systematic semianalytical methodology, supported by a rigorous analytical model while yielding detailed design specifications, forms a reliable tool for the development of novel dual-polarized MG-based devices. In particular, such a design scheme can be highly appealing for applications such as satellite communication systems \cite{afzal2017steering, afzal2021beam} and future cellular networks \cite{jiang2016metamaterial, li2019compact}.

%

\section{Theory}
\label{sec:Theory}
\subsection{Formulation}
\label{subsec:Formulation}
We consider first the structure depicted in Fig. \ref{fig:configuration_fig}, where columns of dog-bone elements are positioned on $z=-h$, parallel to the $\widehat{xy}$ plane, surrounded by a homogeneous medium with permittivity $\varepsilon$ and permeability $\mu$. Harmonic time dependency $e^{j\omega t}$ is assumed and suppressed, defining the wave number $k=\omega\sqrt{\varepsilon\mu}$, the wavelength $\lambda=2\pi/k$, and the operating frequency $f=\omega/\left(2\pi\right)$; the wave impedance is $\eta=\sqrt{\mu/\varepsilon}$.
The printed dog-bone elements have a trace width of $w$ and thickness $t$. They are closely spaced along the $x$ axis in an $l_x$-periodic arrangement, while positioned in a sparse $\Lambda$-periodic formation along $y$. The structure is excited by a TM-polarized plane wave ($E_x=H_y=H_z=0$) incoming from below; once impinging upon the dog-bone elements, it induces electric dipoles directed parallel to the $y$ axis\footnote{While the induced current profile may be more involved \cite{baladi2021equivalent}, we assume sufficiently small dog-bones such that the leading (electric dipole) term in the multipole expansion is sufficient to obtain an accurate description of the scattering phenomena.}, effectively forming a dipole line (secondary) source \cite{felsenbook}.
As $l_x\ll\lambda \Rightarrow \partial/\partial x \approx 0$, we view the $n$th dipole line as a continuous entity (MA) described by an average current density $\vec{J}_n\left(y,z\right)$, related to the dipole moment per unit length of the dog-bones $\frac{p}{l_x}$ via $\vec{J}_n\left(y,z\right)=j\omega\frac{p}{l_x}\delta \left( y-n\Lambda \right )\delta \left( z+h \right )\hat{y}$, where $y_n=n\Lambda$ is the position of the $n$th dipole line.

Correspondingly, for given induced dipole moments, the secondary fields generated by these TM MAs in the described periodic arrangement would follow the solution of the canonical dipole line problem \cite{felsenbook, clemmow2013plane}, reading
\begin{equation}
\label{eq:h_wire1_general}
\begin{aligned}
E_y^{\mathrm{MG}}\left(\vec{r}\right)
    &=
    j\frac{k\omega\eta}{4}
    \frac{p}{l_x}
    \sum_{n=-\infty}^{\infty}
    \frac{\partial^2}{\partial\left(kz\right)^2}
    \left[ H_0^{(2)}\left(k|\vec{r}-\vec{r'_n}|\right)\right] \\
H_x^{\mathrm{MG}}\left(\vec{r}\right)
    &=
    -\frac{k\omega}{4}
    \frac{p}{l_x}
    \sum_{n=-\infty}^{\infty}
    \frac{\partial}{\partial\left(kz\right)}
    \left[ H_0^{(2)}\left(k|\vec{r}-\vec{r'_n}|\right)\right]
\end{aligned}
\end{equation}
where $H^{(2)}_{\nu}(\Omega)$ is the $\nu$th order Hankel function of the second kind, $\vec{r}$ stands for the observation point, and $\vec{r'_n}=\left(0,n\Lambda,-h\right)$ is the location of the $n$th MA.  
We utilize the Poisson summation formula to find the Floquet-Bloch (FB) representation of the scattered fields in our case\cite{tretyakov2003analytical}.
The tangential fields generated by the TM MGs would therefore read
\begin{equation}
\label{eq:fields_TM_MG}
\begin{aligned}
    E_y^{\mathrm{MG}}\left(y,z\right)
    &=
    -j\frac{\eta c}{2\Lambda}
    \frac{p}{l_x}
    \sum_{m=-\infty}^{\infty}
    \beta_m e^{-jk_{t,m}y}
    e^{-j\beta_m|z+h|} \\
    H_x^{\mathrm{MG}}\left(y,z\right)
    &=
    j\frac{kc}{2\Lambda}
    \frac{p}{l_x}
    \sum_{m=-\infty}^{\infty}
    \mathrm{sgn}\{z+h\} e^{-jk_{t,m}y}
    e^{-j\beta_m|z+h|}
\end{aligned}
\end{equation}
with $ k_{t,m}=2\pi m/\Lambda$ and $\beta_m=\sqrt{k^2-k_{t,m}^2}$ ($\Im\{\beta_m\}\leq 0$) being the $m$th mode transverse and longitudinal wavenumbers, respectively. As expected, the tangential magnetic field changes sign when crossing the electric current source sheet \cite{pozar2011microwave}, manifesting the fact that the scattered waves travel away from the MG. 

\begin{figure}[tbp]
\centering
\includegraphics[width=80mm]{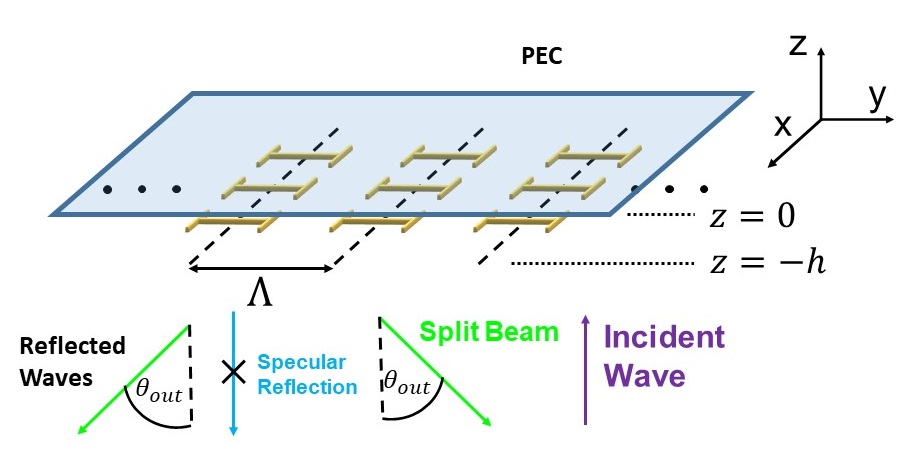}
\caption{Physical configuration of a $\Lambda$-periodic MG for perfect beam splitting of a normally incident TM-polarized plane wave towards $\pm\theta_\mathrm{out}$, featuring the proposed dog-bone elements as MAs.}
\label{fig:TM_beam_splitter_fig}
\end{figure}

Once we have obtained a formulation for the fields scattered off an isolated TM MG, we may proceed and utilize it for deriving analytical expressions for the scattered fields when the MG is embedded in a more involved configuration. In particular, we are interested herein in devising a TM beam splitter, operating in reflect mode. To this end, we 
consider the scenario described in Fig. \ref{fig:TM_beam_splitter_fig}, where the MG is positioned a distance $h$ below a perfect-electric-conductor (PEC) mirror. Similar to previous work \cite{epstein2017unveiling, radi2017metagratings}, to analyze the scattered fields in this scenario we evaluate separately the contributions associated with the external fields (in the absence of the MG), and the secondary fields stemming from the induced dipoles. 

The external fields consist of a normally incident plane wave approaching from $z\rightarrow-\infty$ and its specular reflection from the PEC at $z=0$, namely,
\begin{equation}
\label{eq:fields_external}
\begin{aligned}
    E_y^{\mathrm{ext}}(y,z)
    =
    E_{\mathrm{in}}
    \left(e^{-jkz}-e^{jkz}\right)
    =
    -2jE_{\mathrm{in}}\sin{\left(kz\right)}
\end{aligned}
\end{equation}
The contribution of the secondary fields generated by the (induced) dipole line MG in the presence of the reflecting PEC can be calculated by using \eqref{eq:fields_TM_MG} in conjunction with  
image theory. In the observation region ($z<-h$) below the MG, these are given by 
\begin{equation}
\label{eq:fields_MG_above_PEC}
\begin{aligned}
    &E_y^{\mathrm{PEC-MG}}\left(y,z<-h\right)
    =\\
    &\,\,\,\,\,\,=\frac{\eta c}{\Lambda}
    \frac{p}{l_x}
    \sum_{m=-\infty}^{\infty}
    \beta_m \sin{\left(\beta_mh\right)}
    e^{-jk_{t,m}y}
    e^{j\beta_mz}
\end{aligned}
\end{equation}
The total field will thus be the summation of these two field contributions, namely,
\begin{equation}
\label{eq:fields_total_small}
\begin{aligned}
    E_y^{\mathrm{tot}}(y,z)
    =
    E_y^{\mathrm{PEC-MG}}(y,z)
    +E_y^{\mathrm{ext}}(y,z)
\end{aligned}
\end{equation}
The magnetic field can be subsequently deduced via Maxwell's equations by a simple derivation, yielding, together with \eqref{eq:fields_external}-\eqref{eq:fields_total_small}, explicit expressions for the electric and magnetic tangential fields below the MG ($z<-h$),
\begin{equation}
\label{eq:fields_total}
\begin{aligned}
    E_y^{\mathrm{tot}}
    =
    &-2jE_{\mathrm{in}}\sin{\left(kz\right)} \\
    &+\frac{\eta c}{\Lambda}
    \frac{p}{l_x}
    \sum_{m=-\infty}^{\infty}
    \beta_m \sin{\left(\beta_mh\right)}
    e^{-jk_{t,m}y}
    e^{j\beta_mz}\\
    H_x^{\mathrm{tot}}
    =
    &-2\frac{E_{\mathrm{in}}}{\eta}\cos{\left(kz\right)} \\
    &+\frac{kc}{\Lambda}
    \frac{p}{l_x}
    \sum_{m=-\infty}^{\infty}
    \sin{\left(\beta_mh\right)}
    e^{-jk_{t,m}y}
    e^{j\beta_mz}
\end{aligned}
\end{equation}

By considering these total fields scattered from the overall MG configuration (dog-bone array + PEC), we may now formulate constraints on the various mode amplitudes (coupling coefficients) as to obtain the goal functionality - perfect beam splitting towards $\pm\theta_\mathrm{out}$. In analogy to the TE-polarized case \cite{epstein2017unveiling}, our degrees of freedom in the design would be the period size $\Lambda$; the MA dimensions [specifically, the dog-bone arm length $W$, \textit{cf.} Fig. \ref{fig:configuration_fig}(a)], related to their polarizability; and the distance $h$ between the MG and the PEC reflector. 

\subsection{Perfect beam splitting}
\label{subsec:Propagating mode selection rules}
To synthesize the desired TM beam splitter, we follow the methodology presented in \cite{epstein2017unveiling, radi2017metagratings}, and impose three physical demands on the realized structure. First, we require that all higher-order FB modes ($|m|>1$) would be evanescent, such that real power could only be scattered towards the specular direction (which is always allowed) and to the desired split angles $\pm\theta_\mathrm{out}$ ($m=\pm1$). Correspondingly, according to the FB theorem\cite{bhattacharyya2006phased}, we tune the MG periodicity $\Lambda$ such that the trajectory of the first FB mode would coincide with $\theta_\mathrm{out}$ for a normally incident wave, and demand that all other modes have decaying wave functions. Formally, this translates into 
\begin{equation}
\label{eq:mode_selection}
\begin{aligned}
    &\frac{2\pi}{\Lambda}=k_0\sin{\theta_{\mathrm{out}}}, 
    &\Re\left\{\beta_m \right\}_{|m|\geq 2}=0 \\
    &\Rightarrow \lambda<\Lambda<2\lambda, 
    &\theta_{\mathrm{out}}
    >\ang{30}
\end{aligned}
\end{equation}
This resolves the first degree of freedom, relating the periodicity $\Lambda$ to the output angle $\theta_{\mathrm{out}}$ while identifying the valid angular range.

Second, we demand that no power will be coupled to the specular reflection ($m=0$) mode. This is accomplished by requiring that the field scattered by the MG to this direction would destructively interfere with the contribution of the external fields. Reviewing \eqref{eq:fields_total} and stipulating that the $m=0$ mode amplitude would vanish results in yet another condition
\begin{equation}
\label{eq:first_solution}
\begin{aligned}
    E_{\mathrm{in}}
    \left(\frac{p}{l_x}\right)^{-1}
    =
    \frac{\omega\eta}{\Lambda}
    \sin\left(kh\right)
\end{aligned}
\end{equation}
which indicates the induced dipole strength required for countering the specular reflection of the incident fields.

\begin{figure}[tbp]
\centering
\includegraphics[width=80mm]{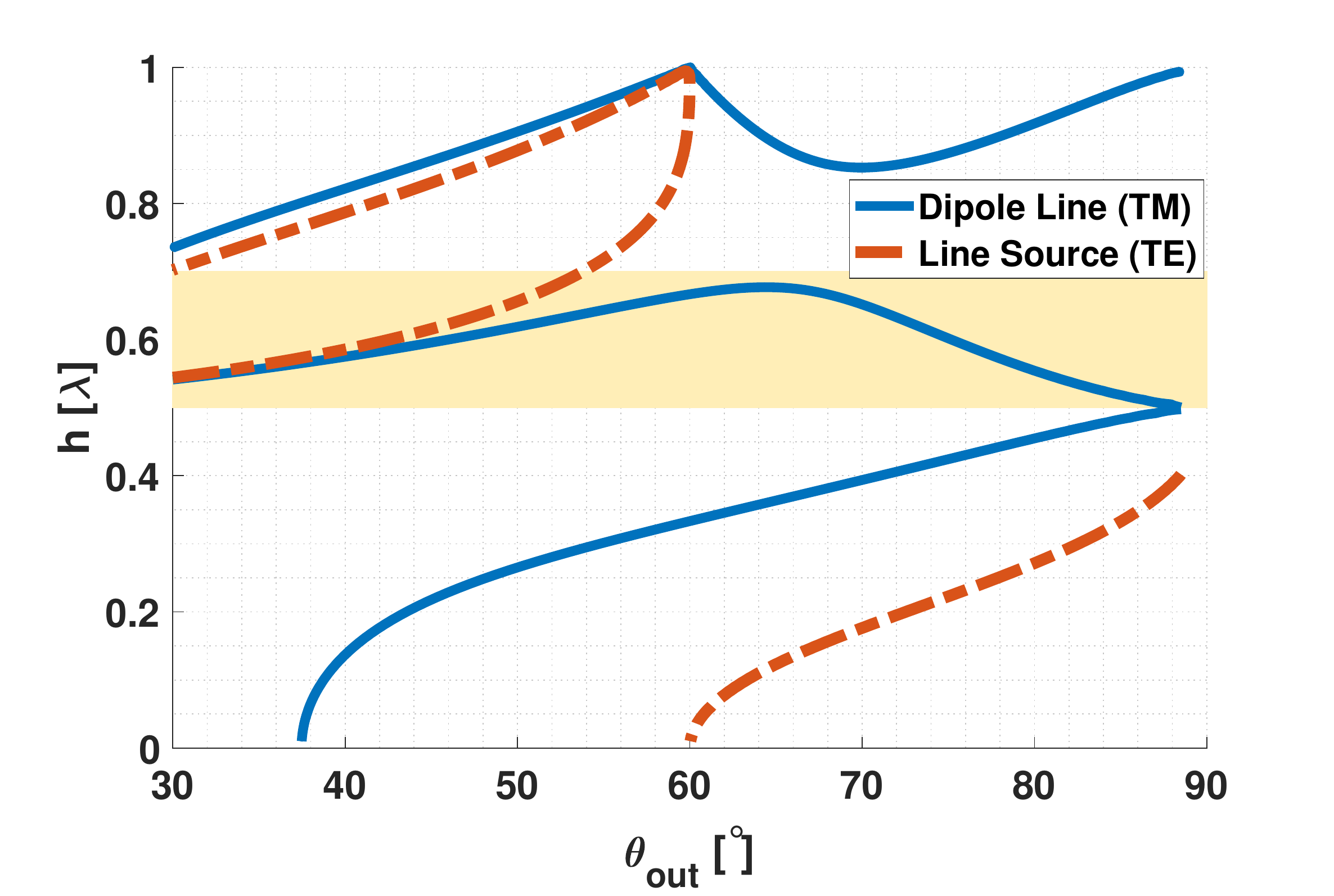}
\caption{Required distance $h$ between the PEC and the MG for facilitating perfect beam splitting, as a function of the split angle $\theta_\mathrm{out}$. The values corresponding to a TM-sensitive design (solid blue, \eqref{eq:powerplane_explicit}) and a TE-sensitive design (dash-dotted orange \cite{epstein2017unveiling}) are presented, where the region highlighted in beige background denotes the solution branches used for the implementation presented in Section \ref{sec:results}.} 
\label{fig:height_fig}
\end{figure}

Lastly, since we wish the designed MG to allow realization via passive and lossless (reactive) components, we demand that power will be overall conserved by the system. Specifically, we require the net real power crossing any given plane $z=z_0<-h$ below the MG would vanish, namely 
\begin{equation}
\label{eq:powerplane}
\begin{aligned}
    P_z^{\mathrm{tot}}
    \left(z=z_0<-h\right)
    =-
    \frac{1}{2}
    \int_{-\Lambda/2}^{\Lambda/2}
    \Re\left\{E_yH_x^*\right\}dy
    =0
\end{aligned}
\end{equation}
Substituting \eqref{eq:fields_total}, \eqref{eq:first_solution}  into \eqref{eq:powerplane} yields, after some algebraic simplifications, the following nonlinear equation
\begin{equation}
\label{eq:powerplane_explicit}
\begin{aligned}
    \sin^2\left(kh\right)
    -2\cos\theta_{\mathrm{out}}
    \sin^2\left(kh\cos\theta_{\mathrm{out}}\right)
    =
    0
\end{aligned}
\end{equation}
the resolution of which identifies the possible dog-bone-PEC spacings $h$ that would enable realization of passive and lossless MG for perfect beam splitting towards $\pm\theta_\mathrm{out}$. 

Figure \ref{fig:height_fig} presents the relation between these distances $h$ and the output angle $\theta_{\mathrm{out}}$, obtained by graphically solving\footnote{Since \eqref{eq:powerplane_explicit} is nonlinear, several solution branches exist in Fig. \ref{fig:height_fig}.} \eqref{eq:powerplane_explicit}; for comparison, the analogous relation corresponding to the TE MG beam splitter presented in \cite{epstein2017unveiling}, based on capacitively loaded wires, is shown as well. Examining the nonlinear equation applicable for the TE-polarized case [Eq. (14) of \cite{epstein2017unveiling}] with respect to \eqref{eq:powerplane_explicit}, one can pinpoint the difference between the two trends as stemming from the different wave impedances associated with TE- and TM-polarized waves \cite{pozar2011microwave}.
We further note that \eqref{eq:powerplane_explicit} is consistent with the results presented in \cite{reconfigurable_strips2018} for a TM MG based on wide metallic strips as meta-atoms. Nonetheless, as will be shown in Section \ref{sec:results}, the MG realization proposed herein for TM-polarized applications is advantageous since it does not interact with TE-polarized waves, while at the same time enabling convenient assessment of the MA geometry.

\subsection{Meta-atom polarizability}
\label{subsec:TM MG Analysis}
Once the position of the MG below the PEC mirror is set via \eqref{eq:powerplane_explicit}, we proceed to finding the suitable MA polarizability of the dog-bones to be placed there, guaranteeing implementation of the perfect beam splitting. Instead of the loaded wires utilized before for realizing TE-polarized MG \cite{epstein2017unveiling} that could sustain a continuous current along them, the interaction with the elementary scatterers used herein are more appropriately described via induced ($y$-directed) dipole moments $p$ [Fig. \ref{fig:configuration_fig}(b)]. Correspondingly, the relation between the local field applied at the dog-bone position to the excited dipole moment is given by the scatterer polarizability \cite{tretyakov2003analytical}
\begin{equation}
\label{eq:polarizability_definition}
\begin{aligned}
    \frac{p}{l_x}
    =
    \frac{\alpha}{l_x}
    E_y^{\mathrm{loc}}\left(y=0,z=-h\right)
\end{aligned}
\end{equation}
where $E_y^{\mathrm{loc}}\left(y=0,z=-h\right)$ stands for the local field at the dipole location, created by all the sources in the given geometry but the one under test, and $\frac{\alpha}{l_x}$ corresponds to the MA polarizability per unit length. We should emphasize that the relation that ties the induced current and the acting fields used herein differs from the one employed in \cite{epstein2017unveiling} in the sense that the acting fields herein exclude the fields generated by the induced dipole moment itself, in consistency with the polarizability concept \cite{tretyakov2000line, tretyakov2003analytical, pulido2018analytical, reconfigurable_strips2018}.

Considering the geometry in Fig. \ref{fig:TM_beam_splitter_fig}, the local field at the vicinity of the reference MA at $(y,z)=(0,-h)$ is composed of three field contributions
\begin{equation}
\label{eq:local_field_MG}
\begin{aligned}
    &E_y^{\mathrm{loc}}\left(y=0,z=-h\right) 
    = \\
    &\,\,\,\,\left[E_y^{\mathrm{ext}}\left(y,z\right) +
    E_y^{\mathrm{intra}}\left(y,z\right) +
    E_y^{\mathrm{image}}\left(y,z\right) 
    \right]_{\left(y,z\right)=\left(0,-h\right)}
\end{aligned}
\end{equation}
where $E_y^{\mathrm{ext}}$ is the external field defined in \eqref{eq:fields_external}, $E_y^{\mathrm{intra}}$ is the intralayer field created by the other induced dipoles residing on the MG plane, and $E_y^{\mathrm{image}}$ stands for the contribution of the image sources corresponding to the latter, formed by the reflecting PEC at $z=0$. Overall, utilizing \eqref{eq:h_wire1_general}-\eqref{eq:fields_external}, the local field can be explicitly formulated as
\begin{equation}
\label{eq:local_field_MG_explicit}
\begin{aligned}
    E_y^{\mathrm{loc}}&\left(y=0,z=-h\right) 
    =
    \\
    =
    2j&E_{\mathrm{in}}\sin{\left(kh\right)}
    -j\frac{\eta\omega}{2}
    \frac{p}{l_x}
    \sum_{n=1}^{\infty}
    \frac{H_1^{\left(2\right)}\left(nk\Lambda\right)}{n\Lambda}
    \\
    &+j\frac{\eta c}{2\Lambda}
    \frac{p}{l_x}
    \sum_{m=-\infty}^{\infty}
    \beta_m
    e^{-2j\beta_mh}
\end{aligned}
\end{equation}
where the intralayer coupling is calculated directly from \eqref{eq:h_wire1_general}. Noting that for large distances $kr\gg1$, 
$|\frac{H_1^{(2)}\left(kr\right)}{kr}|\sim \left(kr\right)^{-3/2}$ attenuates rather quickly \cite{abramowitz1988handbook}, 
we find that the second term in \eqref{eq:local_field_MG_explicit} readily converges as a sum of non-radiative waves (reactive near fields). Obeserving that the FB series in the third term 
includes only a finite number of propagating harmonics (the rest are rapidly decaying evanescent components), provides the necessary guarantee that the entire RHS of \eqref{eq:local_field_MG_explicit} properly converges. 

By substituting \eqref{eq:first_solution}, \eqref{eq:local_field_MG_explicit} back to the definition \eqref{eq:polarizability_definition}, we find that the MA polarizability required for realizing perfect beam splitting of a normally incident TM-polarized wave towards $\pm\theta_\mathrm{out}$ is
\begin{equation}
\label{eq:interraction_const_splitter_explicit}
\begin{aligned}
    \frac{\alpha}{l_x} = 
    \left(
    2j\frac{\omega\eta}{\Lambda}
    \sin^2\left(kh\right)
    -j\frac{\eta\omega}{2}
    \sum_{n=1}^{\infty}
    \frac{H_1^{\left(2\right)}\left(nk\Lambda\right)}{n\Lambda}
    \right. \\
    \quad \left. {}+
    j\frac{\eta c}{2\Lambda}
    \sum_{m=-\infty}^{\infty}
    \beta_m
    e^{-2j\beta_mh}
    \right)^{-1}
\end{aligned}
\end{equation}

This concludes the principal design procedure for the proposed TM-polarized MG. For given operating frequency and split angle $\theta_\mathrm{out}$, we first use \eqref{eq:powerplane_explicit} to extract the PEC-MG separation $h$ required to guarantee complete suppression of the specular reflection via a passive and lossless MG (if proper dipoles are induced on the MAs by the impinging fields). Next, the extracted $h$ is substituted into \eqref{eq:interraction_const_splitter_explicit} to resolve the dipole line polarizability per unit length $\alpha/l_x$ that would ensure that the desired dipole moment would indeed be induced. With these in hand, the overall synthesis of the MG can be completed by setting the MA dimensions to realize the evaluated $\alpha/l_x$, which will be addressed and demonstrated in next section.

\section{Results and discussion}
\label{sec:results}

\subsection{TM-polarized MG beam splitter}
\label{subsec:TM_MG_Synthesis}
To verify and demonstrate the theory developed in Section \ref{sec:Theory}, we utilize it to devise a TM-polarized beam splitter at $f=$20 GHz.
In particular, we aim at devising a set of beam splitters for various prescribed split angles, covering the range between $\ang{35}$ and $\ang{80}$ [complying with \eqref{eq:mode_selection}].
Our first step would be to match a suitable MA geometry to fulfill our requirement for the MA polarizability (see Section \ref{subsec:TM MG Analysis}).
To this end, we establish a method to evaluate the polarizabilty of a given scatterer geometry, following the procedure proposed in \cite{levy2019rigorous,popov2019designing}.

As described in Section \ref{subsec:Formulation}, we focus on dog-bone arrays as meta-atoms for the TM-polarized device (Fig. \ref{fig:configuration_fig}), with their trace width and thickness fixed to $w=3\,\mathrm{mil}\approx 76\mu $m and $t=35\mu$m, respectively, following standard fabrication capabilities. We set their side-arm length to $L=70\,\mathrm{mil}\approx 1.78\mathrm{ mm}$, acting as a constant inductive load, while using the dipole length $W$ to tune the meta-atom response (modify the working point around the resonance).
Correspondingly, we consider a range of values for $W$, serving as the degree of freedom for realizing the desired polarizability values found as per \eqref{eq:interraction_const_splitter_explicit}.
Subsequently, to characterize the TM MG polarizability corresponding to a given $W$, we place the MA in a $\Lambda$-periodic formation in a full-wave solver (CST Microwave Studio), excite it with a normally incident plane wave [as shown in Fig. \ref{fig:configuration_fig} (b)], and record the reflection coefficient of the fundamental mode $R_0$.
Since this coefficient can be readily related to the induced dipole by considering the $m=0$ term in \eqref{eq:fields_TM_MG}, namely, $R_0=-\frac{j\omega\eta}{2\Lambda E_{\mathrm{in}}}\frac{p}{l_x}$, we can use the recorded $R_0$ in \eqref{eq:polarizability_definition} to estimate the effective polarizability of this specific dipole line geometry, reading
\begin{equation}
\label{eq:polarizability_extraction_singleMG}
\begin{aligned}
    &\frac{\alpha}{l_x}
    =
    \left(
    -j\frac{\eta\omega}{2\Lambda}
    \frac{1}{R_0}
    -j\frac{\eta\omega}{2}
    \sum_{n=1}^{\infty}
    \frac{H_1^{\left(2\right)}\left(nk\Lambda\right)}{n\Lambda}
    \right)^{-1}
\end{aligned}
\end{equation}

\begin{figure}[tbp]
\centering
\includegraphics[width=80mm]{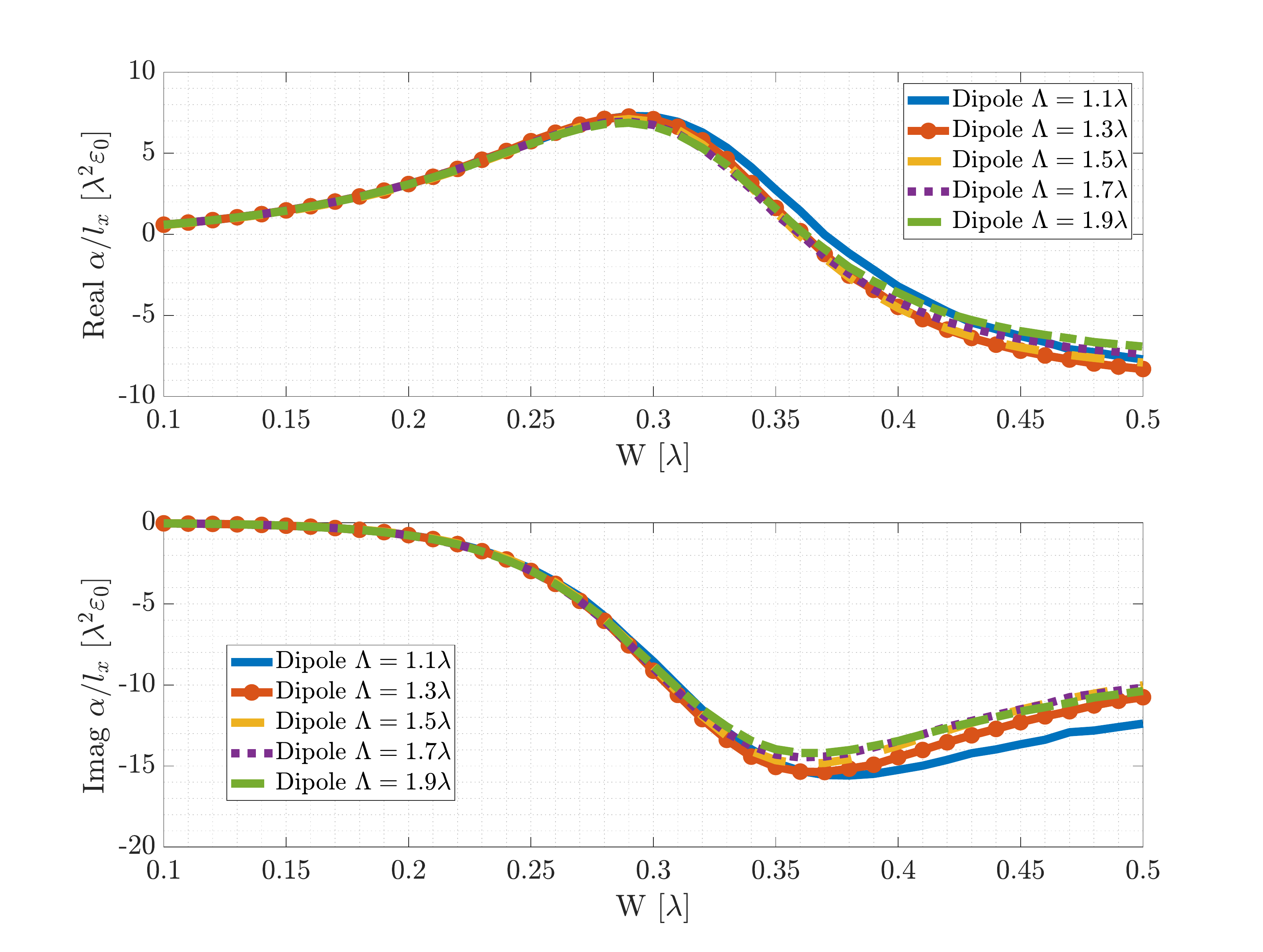}
\caption{Effective dog-bone array polarizability as a function of the dipole length [Fig. \ref{fig:configuration_fig}(a)], for various MG periodicities $\Lambda = 1.1\lambda$ (solid blue), $\Lambda = 1.3\lambda$ (orange with circle markers), $\Lambda = 1.5\lambda$ (dashed yellow), $\Lambda = 1.7\lambda$ (dotted purple), and $\Lambda = 1.9\lambda$ (dash-dotted green). Top and bottom plots correspond to real and imaginary parts of the polarizability $\alpha/l_x$, respectively, estimated following Section \ref{subsec:TM_MG_Synthesis}. The curve corresponding to $\Lambda=1.5\lambda$ is selected as the basis for converting a given polarizability requirement into a realistic trace geometry, as part of the realizations considered in Sections \ref{subsec:TM_MG_Synthesis} and \ref{subsec:Dual Polarized Beam Splitter Synthesis}. 
}
\label{fig:polarizability_fig}
\end{figure}

Figure \ref{fig:polarizability_fig} presents the extracted effective polarizabilities of the TM-susceptible dog-bone MAs as a function of the dipole length, for a series of period sizes $\lambda<\Lambda<2\lambda$.
The plots confirm that the polarizability is an inherent property of the dog-bone column alone, almost independent of the period size. The slight differences found in this regard for different values of $\Lambda$ are attributed to approximated dipole model used in our formulation leading to \eqref{eq:polarizability_extraction_singleMG}, neglecting high order multipoles associated with the dog-bone side-arms, which may affect the mutual coupling between adjacent MAs. Nonetheless, since the differences between the curves is very small, we choose to use the one corresponding to $\Lambda = 1.5\lambda$ for devising the MG realizations herein and in subsequent sections.

It should be noted that the polarizability trends in Fig. \ref{fig:polarizability_fig} match the polarizability of an inductively loaded dipole (loaded wire \cite{tretyakov2000line, tretyakov2003analytical}), as previously reported in the literature \cite{pulido2018analytical}. Compared to pristine dipoles, it can be seen that the dog-bone side-arms, acting as inductive loads, shift the resonance frequencies down, which manifests itself as a shift of the polarizability graph towards smaller $W$. 
Similarly, since the polarizability is proportional to the electrical length of the dipole (and not its absolute length), correspondence can be found also with the typical frequency dependency of dipole polarizability \cite{pulido2018analytical}.

\begin{figure}[htb]
\centering
\includegraphics[width=80mm]{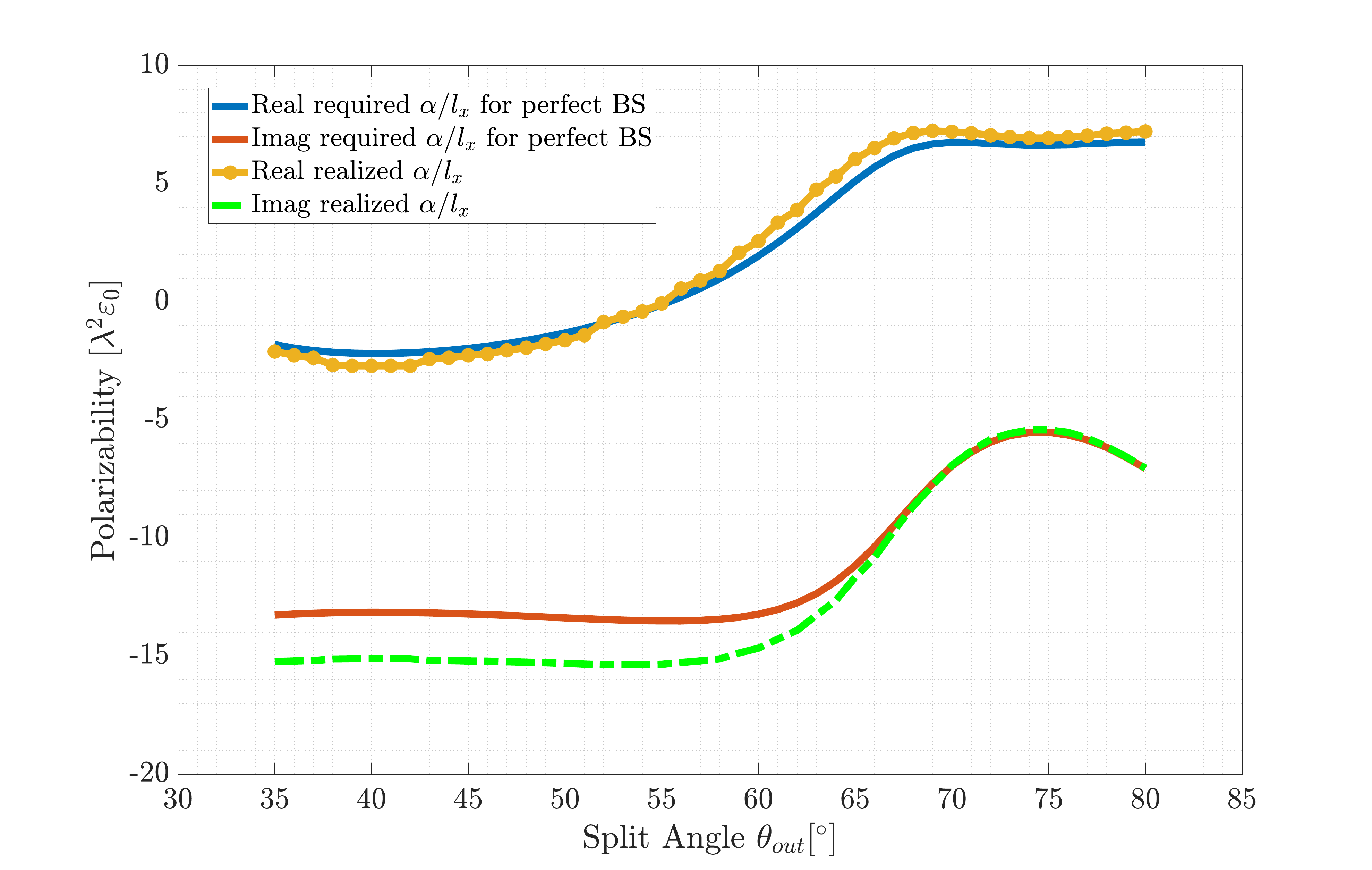}
\caption{TM MG beam splitters ($f=20$ GHz) required polarizability real and imaginary parts, with their actual matched polarizability from the lookup table (Fig. \ref{fig:polarizability_fig}).}
\label{fig:match_dogbone}
\end{figure}

With the devised dipole lines in our hands, we continue and implement the theory as mentioned in Section \ref{subsec:Formulation}. Our degrees of freedom in the design would be the period size ($\Lambda$), the dog-bone (meta-atom) arm length ($W$), and the MG separation distance from the reflector ($h$). For a given $\theta_\mathrm{out}$, we (i) use the FB theorem \eqref{eq:mode_selection} to set the MG periodicity; (ii) enforce specular reflection elimination and power conservation by utilizing a spacing $h$ which solves \eqref{eq:powerplane_explicit} for the prescribed $\theta_\mathrm{out}$ (we focus on the solution branch highlighted in Fig. \ref{fig:height_fig}, which covers the entire angular splitting range); and (iii) calculate the required meta-atom polarizability $\alpha$ for the chosen $h$ using \eqref{eq:interraction_const_splitter_explicit}, translating it into the actual dog-bone dimensions by finding the closest geometry in our lookup table (Fig. \ref{fig:polarizability_fig}) that provides the goal poarizability.

We follow this approach for the range of angles $\theta_\mathrm{out}\in[35^\circ, 80^\circ]$. The corresponding 
required $\alpha/l_x$ (solid lines) are presented in Fig. \ref{fig:match_dogbone} for each of the split angles, along the closest achievable value (by means of least square error) using the specific dog-bone geometry (dot-dashed or circle markers) as deduced from Fig. \ref{fig:polarizability_fig}. As can be seen, the chosen MA geometry satisfactorily allows meeting the demands in terms of polarizability, with a slight deviation in the imaginary part for split angles below $60^\circ$. Finally, the prescribe polarizability (Fig. \ref{fig:match_dogbone}) is translated into actual dog-bone dimensions (dipole length $W$) with the aid of Fig. \ref{fig:polarizability_fig}, setting the last degree of freedom and thus finalizing the design. The resultant beam splitters were then defined and simulated in a full-wave solver (CST Microwave Studio), for verification. The final design parameters along with the recorded TM MG beam splitter performance are presented in Table \ref{tab:metagrating_performance_10GHz}.

\begin{table*}[t]
\centering
\begin{threeparttable}[b]
\renewcommand{\arraystretch}{1.3}
\caption{Design specifications and simulated performance of the designed metagratings operating at $f=20\mathrm{GHz}$ (corresponding to Fig. \ref{fig:TM_beam_splitter_fig}).}
\label{tab:metagrating_performance_10GHz}
\centering
\begin{tabular}{l|c|c|c|c|c|c|c|c|c|l}
\hline \hline
$\theta_\mathrm{out}$ 
& $35^\circ$ & $40^\circ$ & $50^\circ$ 
& $60^\circ$ & $70^\circ$ & $80^\circ$ \\ 
\hline \hline \\[-1.3em]
	\begin{tabular}{l} $\Lambda [\lambda]$ \end{tabular}
	 & $1.74$ & $1.55$ & $1.3$ 
	& $1.15$ & $1.06$ & $1.015$  \\	\hline	
	 \begin{tabular}{l} $h [\lambda]$ \end{tabular}
	 & $0.556$ & $0.575$ & $0.618$ 
	& $0.667$ & $0.651$ & $0.554$  \\	\hline
	 \begin{tabular}{l} $W [\lambda]$ \end{tabular}
	 & $0.378$ & $0.380$ & $0.365$ 
	& $0.340$ & $0.288$ & $0.299$  \\	\hline 	  
	 \begin{tabular}{l} Splitting Efficiency \end{tabular}
	 & $93.8\%$ & $94.6\%$ & $98.4\%$ 
	& $98.8\%$ & $98.0\%$ & $90.8\%$  \\	\hline 
	\begin{tabular}{l} Specular reflection \end{tabular}
	 & $0.18\%$ & $0.79\%$ & $0.017\%$ 
	& $0.22\%$ & $0.69\%$ & $0.97\%$   \\	\hline 
	 \begin{tabular}{l} Losses \end{tabular}
	 & $6.02\%$ & $4.61\%$ & $1.58\%$ 
	& $0.98\%$ & $1.31\%$ & $8.23\%$  \\	\hline
	\begin{tabular}{l} Bandwidth \end{tabular}
	 & $3\%$ & $4.73\%$ & $13.6\%$ 
	& $21.75\%$ & $19.85\%$ & $26.00\%$  \\		
\hline \hline
\end{tabular}
\end{threeparttable}
\end{table*}

\begin{figure}[htb]
\centering
\includegraphics[width=79mm]{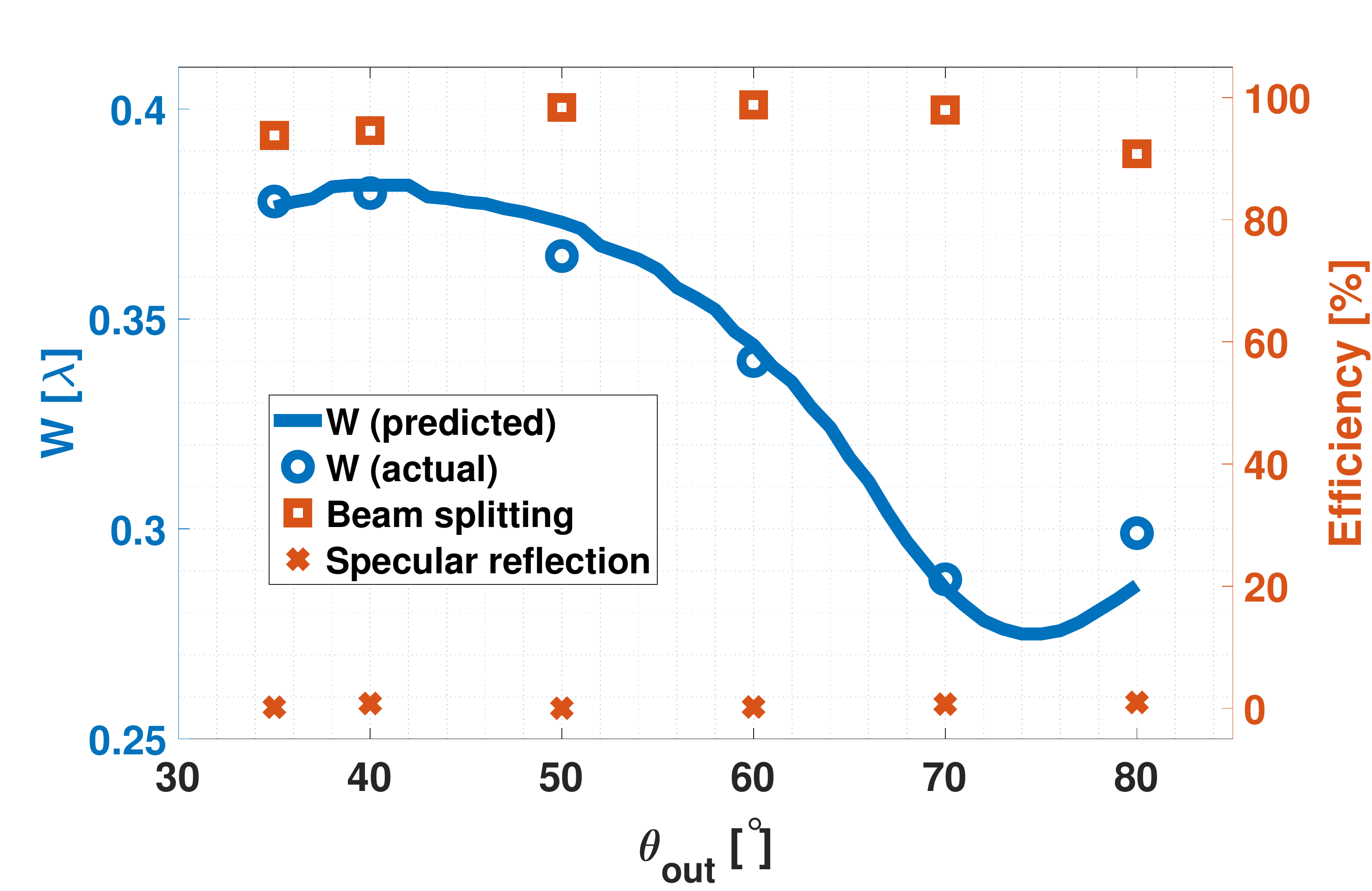}
\caption{Designed TM MG beam splitters ($f=20$ GHz). Semianalytically predicted dog-bone widths as a function of goal split angle (solid blue) are compared to the optimal $W$ obtained from CST (circles), along with simulated beam-splitting (red squares) and specular reflection ($\times$) efficiencies.}
\label{fig:Efficiency_fig}
\end{figure}

Figure \ref{fig:Efficiency_fig} shows the dog-bone dipole length $W$ as a function of the desired split angle as predicted by the described synthesis method (solid blue), along with the actual $W$ values (blue circles) that yield highest beam splitting efficiency as obtained from limited parametric sweeps in CST around the predicted value.
As shown by these plots, corresponding to the left $y$ axis of Fig. \ref{fig:Efficiency_fig}, the results match very well, indicating that our seimanalytical design method achieves a near optimal design for every split angle, without actually involving full-wave optimizations. The orange markers corresponding to the right $y$ axis of Fig. \ref{fig:Efficiency_fig} verify that complete specular reflection suppression is indeed obtained for all considered angles (x markers), while the beam-splitting efficiency approaches unity (squares). These conclusions are confirmed quantitatively by the values presented in Table \ref{tab:metagrating_performance_10GHz}.

A closer examination of Table \ref{tab:metagrating_performance_10GHz} reveals that the fraction of power coupled to the specular reflection and the $\pm 1$ modes ($\theta_\mathrm{out}$ directed beams) does not sum up to unity at all points; this is due to inevitable conductor loss in the realistic copper traces realizing the dog-bone in practice, which is, however, quite minor. Similar observations has been made for TE-polarized MGs \cite{epstein2017unveiling}, where the copper traces implementing the beam splitter absorb a fraction of the power. 


\begin{figure}[htb]
\centering
\includegraphics[width=79mm]{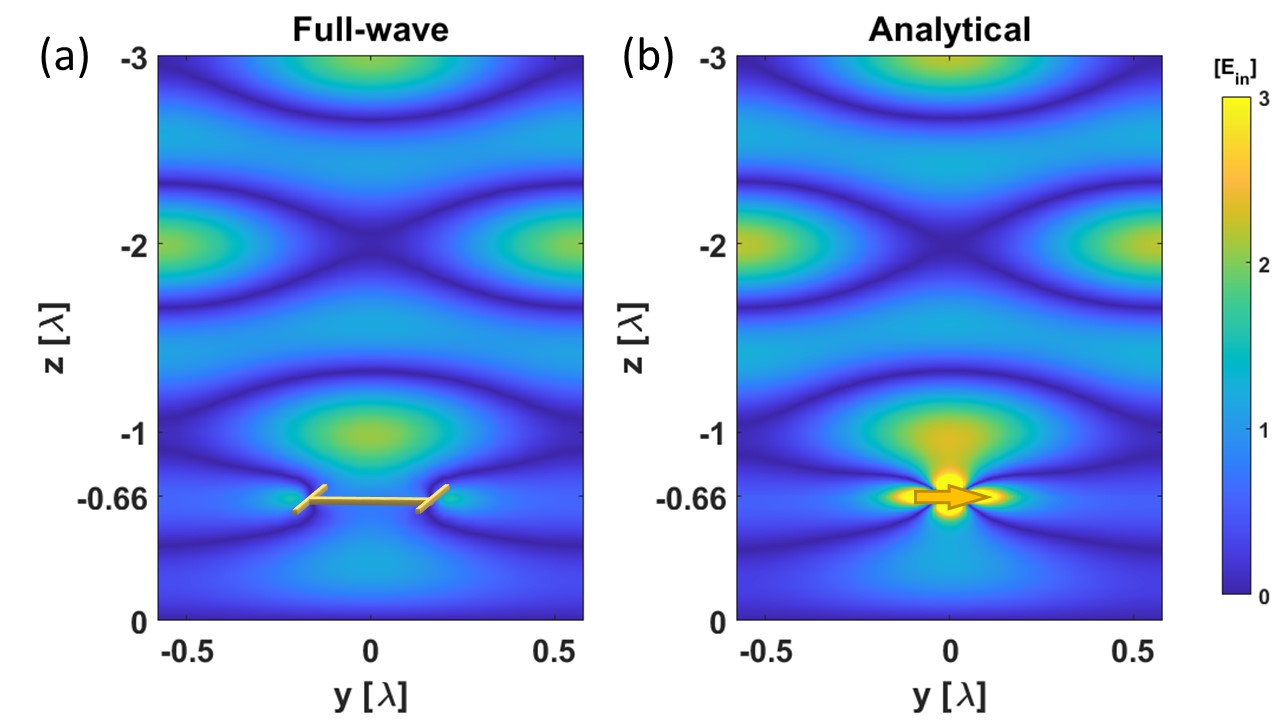}
\caption{Electric-field distributions $|\operatorname{Re}\{E_y\left(y,z\right)\}|$ for TM beam splitting MGs (fig. \ref{fig:TM_beam_splitter_fig}) operating at $20$ GHz, excited from below by an incident plane wave. A single period $\Lambda$ (eq. \eqref{eq:mode_selection}) is shown for a split angle $\theta_\mathrm{out}=60^\circ$. The MGs is placed on the $z=-0.66$ plane correspondingly with eq. \eqref{eq:powerplane_explicit} and fig. \ref{fig:height_fig}. (a) full-wave simulation result. (b) Analytical prediction result. }
\label{fig:fields_comparison}
\end{figure}

To further probe the fidelity of our design method, we compare in Fig. \ref{fig:fields_comparison} the analytically predicted electric fields \eqref{eq:fields_total} with the fields recorded by the full-wave solver when simulating the actual dog-bone configuration specified in Table \ref{tab:metagrating_performance_10GHz}, for a representative case (MG realizing beam splitting towards $\theta_\mathrm{out}=\ang{60}$).
As can be clearly observed, the two field snapshots agree very well, except in the regions that are 
very close to the dog-bone construct. This can be explained by the finite size of the dog-bone element, which is not accounted for in the theoretical model, which analyzes it as an infinitesimal $y$-directed dipole; a similar observation has been made for TE-polarized MGs \cite{epstein2017unveiling}.

These results confirm that the formulated semianalytical scheme and chosen MA realization indeed enable systematic design of TM-polarized MGs, demonstrating efficient beam manipulation with a sparse PCB-compatible design while avoiding full-wave optimization.

\subsection{Dual-polarized beam splitter}
\label{subsec:Dual Polarized Beam Splitter Synthesis}

\begin{figure}[tbh]
\centering
\includegraphics[width=79mm]{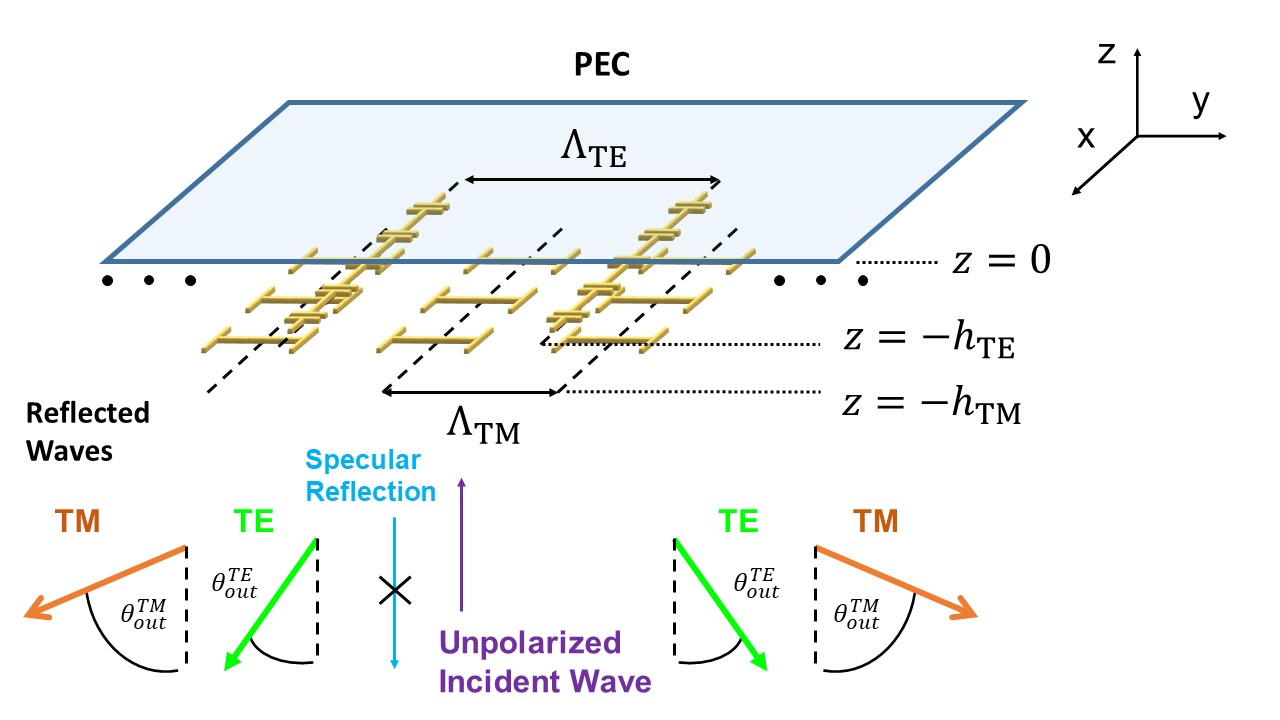}
\caption{Dual-polarized MG beam splitter configuration. $\Lambda_{\mathrm{TM}}$-periodic TM MG are interleaved with $\Lambda_{\mathrm{TE}}$-periodic TE MG, positioned $h_{\mathrm{TM}}$ and $h_{\mathrm{TE}}$ respectively below a PEC mirror.}
\label{fig:dp_figure}
\end{figure}

\begin{figure*}[!hbt]
\centering
\includegraphics[width=180mm]{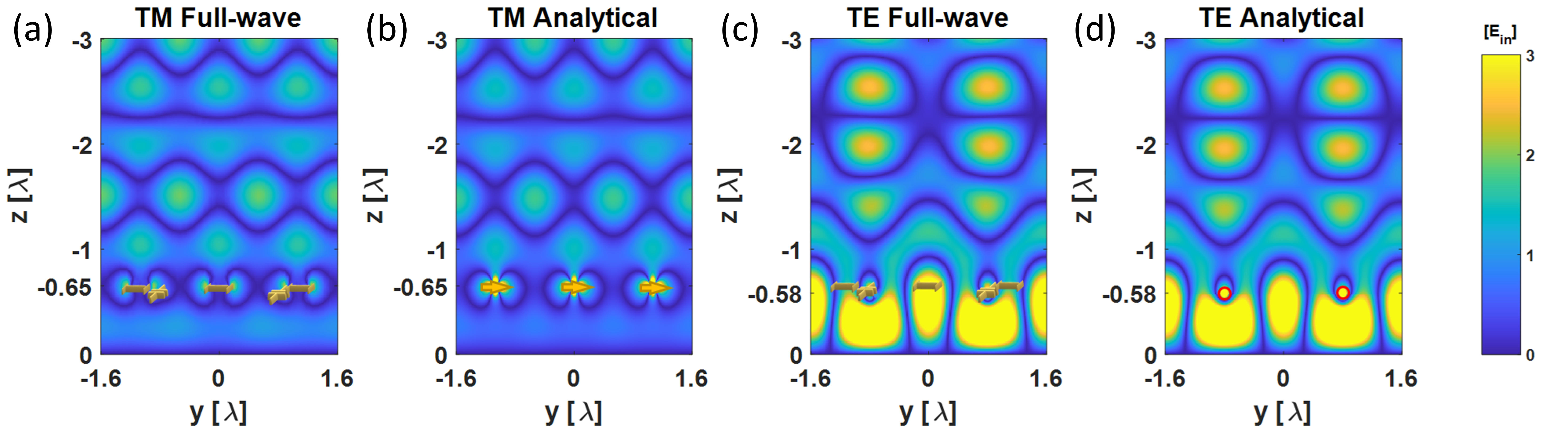}
\caption{Electric-field distributions for dual-polarized beam splitting MGs (Fig. \ref{fig:dp_figure}) operating at $20$ GHz, excited from below by an incident plane wave. A single macro-period $\Lambda_\mathrm{tot}=2\Lambda_{\mathrm{TE}}=3\Lambda_{\mathrm{TM}}$ with the split angles $\theta_\mathrm{out}^\mathrm{TM}=70^\circ$ and $\theta_\mathrm{out}^\mathrm{TE}=38.79^\circ$ shown. The two MGs are placed distances of $h_\mathrm{TE}=0.575\lambda$ and $h_\mathrm{TM}=0.651\lambda$ from the reflector for the TE and TM polarizations respectively. (a) full-wave simulation result for TM response $|\operatorname{Re}\{E_y\left(y,z\right)\}|$. (b) Analytical prediction for TM response $|\operatorname{Re}\{E_y\left(y,z\right)\}|$. (c) full-wave simulation result for TE response $|\operatorname{Re}\{E_x\left(y,z\right)\}|$. (d) Analytical prediction for TE response $|\operatorname{Re}\{E_x\left(y,z\right)\}|$.}
\label{fig:dp_figure_fields}
\end{figure*}

As discussed in Section \ref{Introduction}, one of the main advantages of the proposed PCB-compatible TM MG geometry is that it does not interact with TE-polarized fields (the dominant induced dipoles are along the $y$ axis, with a negligible polarizability component along the $x$ axis \cite{capolino2012equivalent,cui2020dual}).
Consequently, one can readily devise dual-polarized MGs with independent response to TE and TM polarized fields by combining the TM MG configuration presented herein with the TE MG configuration reported in \cite{epstein2017unveiling} (based on loaded wires along $x$, which interacts mostly with TE-polarized fields) on the same PCB. Since either design is practically sensitive to only a single polarization, one can simply follow separately the design schemes for the TM (this work) and TE \cite{epstein2017unveiling} polarized MGs, and then define the two resultant copper traces together on the same board.

We demonstrate this by synthesizing a dual-polarized MG beam splitter, reflecting different polarization components into different angles. In particular, we use the scheme developed herein to design a $\theta_\mathrm{out}^\mathrm{TM}=\ang{70}$ TM beam splitter with a corresponding period size $\Lambda_{\mathrm{TM}}=\lambda/\sin\theta_\mathrm{out}^\mathrm{TM}$, and the one presented in \cite{epstein2017unveiling} to design a $\theta_\mathrm{out}^\mathrm{TE}=\ang{38.79}$ TE beam splitter with a period size $\Lambda_{\mathrm{TE}}=\lambda/\sin\theta_\mathrm{out}^\mathrm{TE}$ (Fig. \ref{fig:dp_figure}). These specific split angles were chosen such that a convenient macro-period could be defined for the combined structure (containing both the TM dog-bone and the TE loaded wire MGs), with $\Lambda_\mathrm{tot}=2\Lambda_{\mathrm{TE}}=3\Lambda_{\mathrm{TM}}$. Correspondingly, according to Fig. \ref{fig:height_fig}, each polarization response requires a different PEC-MG spacing, namely, $h_\mathrm{TE}=0.575\lambda$ and $h_\mathrm{TM}=0.651\lambda$, which indicate the distance for the TE and TM polarizations, respectively. Similarly, the required dog-bone width $W_\mathrm{TM}$ and loaded-wire capacitor width $W_\mathrm{TE}$ can be found by invoking separately the methodologies outlined herein (Table \ref{tab:metagrating_performance_10GHz}) and in \cite{epstein2017unveiling}, respectively.

Subsequently, and without any further optimization, we superimpose the two designs in CST Microwave Studio, forming a single structure with interleaved TE and TM MAs (Fig. \ref{fig:dp_figure}). Running the full-wave solver with the resultant dual polarized MG beam splitter reveals that, as expected, independent beam splitting for the two polarizations has been achieved, with less than $1\%$ of the incident power coupled to undesired specular reflection. The design achieves $90\%$ and $95\%$ beam power splitting efficiency for the TE and TM incident beams respectively. The remaining power dissipates in the copper traces due to the aforementioned conductor loss. This verifies that the two MA configurations indeed do not interfere with one another, effectively separating between the TE- (deflected towards $\theta_\mathrm{out}^\mathrm{TE}=\pm\ang{38.79}$) and TM- (deflected towards $\theta_\mathrm{out}^\mathrm{TM}=\pm\ang{70}$) polarized components of the incident wave. 

The electric fields within the device (Fig. \ref{fig:dp_figure_fields}) further confirm these conclusions, additionally indicating the good correspondence between analytical and full-wave predictions. 
It should be noted that the device performance was found to be insensitive to the lateral offset between the two gratings, providing further evidence to the negligible coupling between the two polarizations in the chosen MAs.

The presented results verify the successful achievement of our goal in this work: proposing a TM-susceptible MG that does not interact with TE-polarized fields, enabling seamless integration of single-polarized MGs into a dual-polarized device with a separable polarization response, without requiring any redesign or optimization.

\section{Conclusion}
To conclude, we have presented a detailed semianalytical approach for designing PCB compatible TM MGs, providing a holistic methodology translating user defined requirements into detailed fabrication-ready devices based on canonical dipole line and particle polarizability models. We have demonstrated successfully how this methodology can be used for implementing a reflective beam splitter, which can be designed to deflect beams towards a wide range of oblique angles with high efficacy. We have further shown that since our proposed (dog-bone based) TM-susceptible MG does not interact with TE fields, it can be utilized to realized dual-polarized MG constructs by seamlessly combining independently devised TE-susceptible (loaded-wire based) MG formations developed in previous work; a polarization-dependent beam splitter was presented based on this idea. These results, verified via full-wave simulations, are expected to lay a rigorous and convenient framework for the development of advanced dual-polarized MG-based devices, establishing a set of effective tools for meeting the requirements of practical communication systems with sparse and rapidly designed composites. 




\end{document}